\documentclass[%
 reprint,
superscriptaddress,
 amsmath,amssymb,
  aps,
]{revtex4-2}

\usepackage{graphicx}
\usepackage{dcolumn}
\usepackage[normalem]{ulem}
\usepackage{bm}

\usepackage{physics}
\usepackage{xcolor}
\usepackage{hyperref}

\usepackage{amsmath,amssymb}

\newcommand*\Laplace{\mathop{}\!\mathbin\bigtriangleup}

\usepackage{siunitx}

\usepackage{amsmath}
\usepackage{amssymb}

\usepackage{soul}

\usepackage{lipsum}
\usepackage{mathtools}

\begin{document}

\title{QCL active region overheat in pulsed mode: effects of non-equilibrium heat dissipation on laser performance}

\author{Ivan I. Vrubel}
 \affiliation{Ioffe Institute, Russian Academy of Sciences, Politekhnicheskaya ul. 26, 194021 St. Petersburg, Russia}
\email{ivanvrubel@ya.ru}

\author{Evgeniia D. Cherotchenko}
\affiliation{Ioffe Institute, Russian Academy of Sciences, Politekhnicheskaya ul. 26, 194021 St. Petersburg, Russia}

\author{Dmitry~A.~Mikhailov}
\affiliation{Ioffe Institute, Russian Academy of Sciences, Politekhnicheskaya ul. 26, 194021 St. Petersburg, Russia}

\author{Dmitry~V.~Chistyakov}
\affiliation{Ioffe Institute, Russian Academy of Sciences, Politekhnicheskaya ul. 26, 194021 St. Petersburg, Russia}

\author{Vladislav V. Dudelev}
\affiliation{Ioffe Institute, Russian Academy of Sciences, Politekhnicheskaya ul. 26, 194021 St. Petersburg, Russia}

\author{Grigorii~S.~Sokolovskii}
\affiliation{Ioffe Institute, Russian Academy of Sciences, Politekhnicheskaya ul. 26, 194021 St. Petersburg, Russia}

\begin{abstract} 
Quantum cascade lasers are of high interest in the scientific community due to unique applications utilizing the emission in mid-IR range. The possible designs of QCL are quite limited and require careful engineering to overcome some crucial disadvantages. One of them is an active region (ARg) overheat, that significantly affects the laser characteristics in the pulsed operation mode. 

In this work we consider the effects related to the non-equilibrium temperature distribution, when thermal resistance formalism is irrelevant. We employ the heat equation and discuss the possible limitations and structural features stemming from the chemical composition of the AR. We show that the presence of alloys in the ARn structure fundamentally limits the heat dissipation in pulsed and CW regimes due to their low thermal conductivity. Also the QCL post-growths affects the thermal properties of a device only in (near)CW mode while it is absolutely invaluable in the pulsed mode.

\end{abstract}

\maketitle

\section{Introduction}
The idea of quantum cascade lasers appeared as early as in 1971\cite{kazarinov1971} and now it is the most promising platform for mid-IR sources and based applications. The last vary from gas-sensing, industrial process control and environmental monitoring technologies to telecommunication and infrared countermeasures. Some of these applications utilize the high-power laser emission which in fact appears to be a tricky challenge. As discussed in \cite{Razeghi15,Mawst2022} in QCL only a small part of input energy is converted into light. The major part is released via heating which in turn lows down the laser performance. Currently wall-plug efficiencies (WPE) for mid-IR QCL at room temperatures in pulsed and cw modes reach 31\% and 22\% respectively\cite{Razeghi3120, Wang20}. The reason for such unimpressive results lies in the internal heating dynamics of a laser. The thermal conductivity of a QCL ARn is essentially anisotropic and very low in the direction perpendicular to the plane of heterostructure layers. 
The lower thermal conductivity is the higher temperature of ARn is achieved during operation CW regime\cite{Wang4120Crio}, which leads to emission degradation and lower WPE. Obviously the same trend is valid for pulsed regime.


The improvement of power and efficiency laser performance involves thorough material engineering\cite{Cherotchenko22, Faugeras05}, and, which is even more important, careful thermal analysis and corresponding structural engineering.  The last includes the study of QCL geometry and ARn conductivity\cite{Evans06}, the effects of cladding and heat-sink form factors, affecting the thermal resistance\cite{Chaparala11}, and design of external cooling systems and its effeciency\cite{Chen06}. In turn the experimental thermal analysis and management techniques are quite complicated and mostly do not allow to study processes inside the AR. General methods, like CCD thermoreflectance\cite{wang22,Pierscinska17,Pierscinski12} scan only the facet which is not informative about the inner processes. A few experiments allow determination of ARn electrical properties\cite{Ladutenko20} from which one can indirectly draw some conclusions about thermal properties.
In this work we discuss the fundamental limitations to the QCL performance arising from the temperature modelling and show that the thermal properties in pulsed regime are insensitive to the laser design, while main features of the laser engineering play the major role in (near)CW mode. 

\section{Results}

\subsection{Experiment}
\begin{figure}
\includegraphics[scale=0.3]{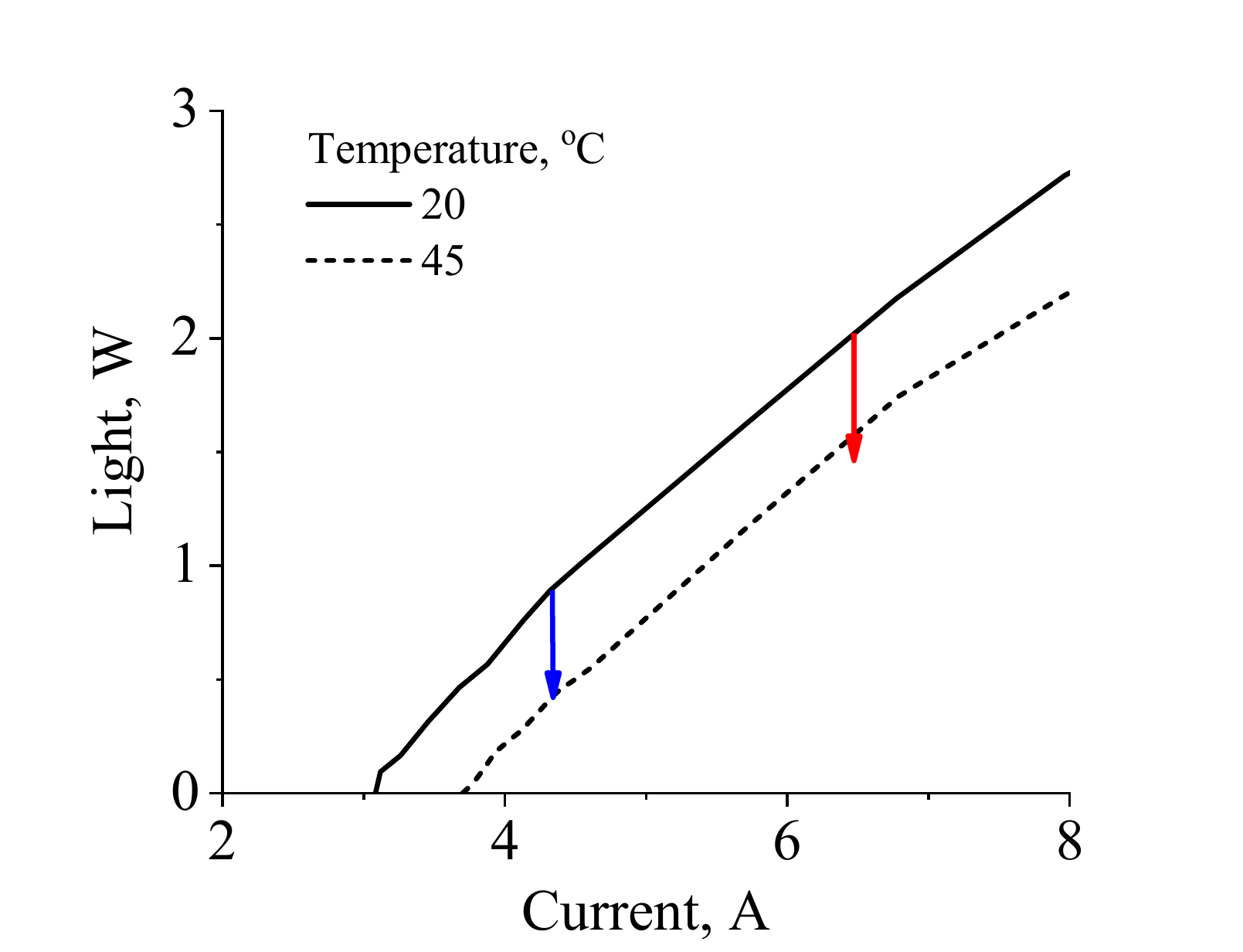}
\includegraphics[scale=0.3]{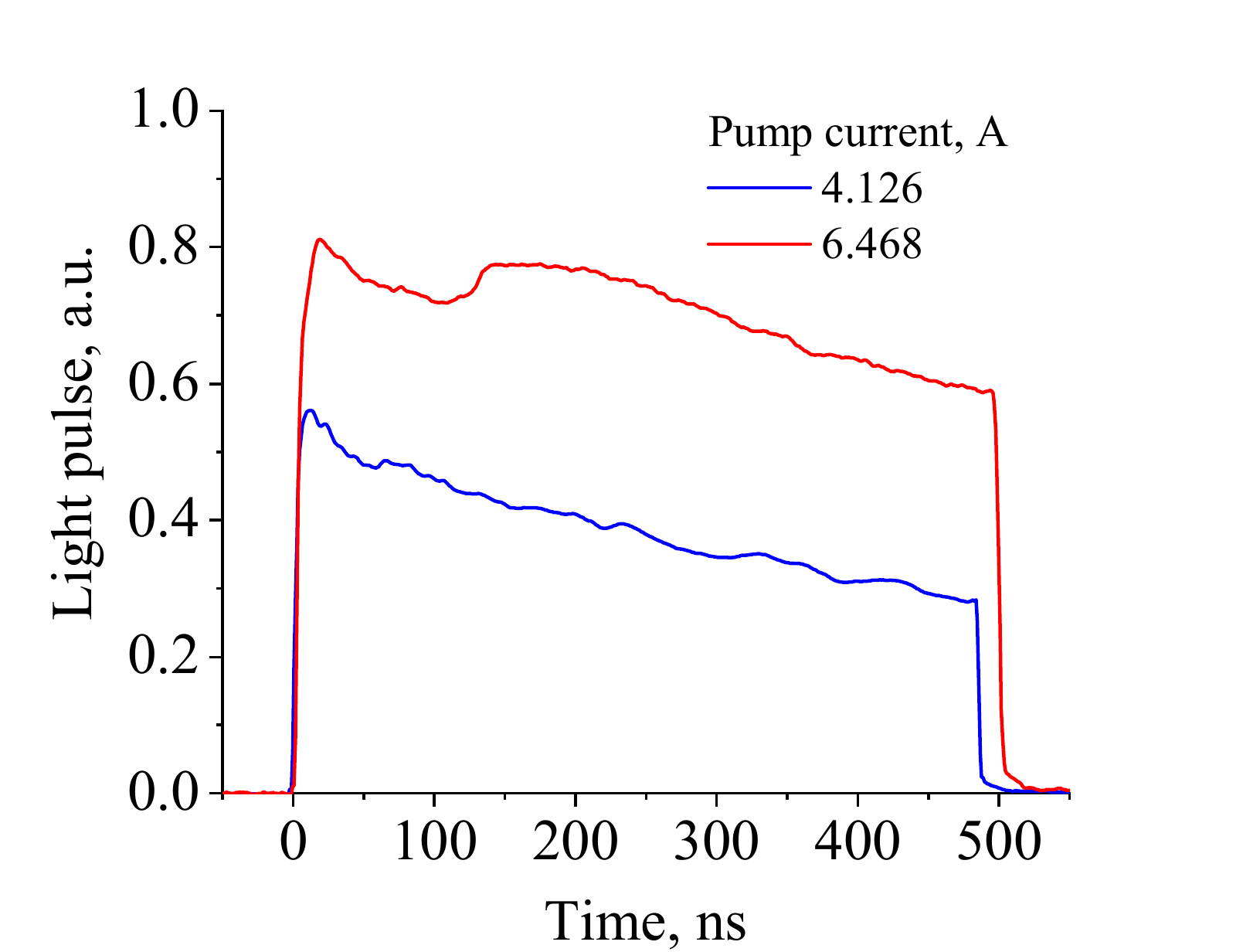}
\caption{Upper panel - regular 
 sample light current characteristics, measured at two ambient temperatures in pulsed regime. Lower panel - time dependent light pulses of a sample pumped by two specified values of pump current. The time-dependent degradation of the light intensity related to the heating is also depicted in the upper panel using vertical arrows with corresponding color.}
\label{fig-exp}
\end{figure}

In this work we used the QCL samples similar to the ones used in Ref.\cite{Cherotchenko22}. The active region comprised a superlattice of {alternating} 
\textbf{In$_{0.53}$Ga$_{0.47}$As}/Al$_{0.48}$In$_{0.52}$As layers that were lattice-matched with the InP substrate~\cite{Babichev17}. The~thicknesses of the sequential layers were
\textbf{{2.4}}/2.4/\textbf{2.6}/2.1/\textbf{2.6}/1.8/\textbf{2.7}/1.6/\textbf{2.9}/1.7/\textbf{3.1}/2.5/\textbf{4.4}/
1.2/\textbf{5.2}/1.2/\textbf{5.3}/1.0/\textbf{1.7}/4.3~nm. The heterostructures were subjected to a post-growth fabrication of QCL chips with 40~$\mu$m stripes and 3~mm cavity lengths. All samples were tested under 500~ns pulsed pumping with a 12~kHz repetition rate.
Here we consider different QCL samples operating under varying current pump pulse at two given temperatures - 20$^0$C and 45$^0$C - of the heat sink. In the very beginning on the first pump pulse the temperature of the ARn is close to the temperature of the thermal reservoir. Under operation the continuously releasing Joule heat results in increase of phonon assisted relaxation rates which in turn leads to degradation of the light intensity with time. In Fig.~\ref{fig-exp} we show the typical light-current characteristics (upper panel) and time-dependent pulse intensities (lower panel) measured in experiment. Red and blue curves at the lower panel show the pulse evolution at two different current amplitudes. Red and blue arrows at the upper panel relate the maximum and minimum light intensity for both values of the pump current. 
20$^0$C and 45$^0$C degrees for the temperature initial conditions of the heat sink were estimated for adiabatic regime, when there is no heat transfer between the ARn and \textcolor{black}{claddings or  any other parts surrounding the structure}. Indeed it appears that the degradation of the light intensity stops in the vicinity of light-current characteristic measured at temperature of 45$^0$C. Thus it is reasonable to deduce that the overheat of a sample ARn is as high as approximately 25~K in the end of 500~ns pulse. In Supplementary materials we show more measurements, where one can find that all the samples exhibit almost the same slope of light pulse degradation.

\subsection{Modeling}

\begin{figure}
\includegraphics[scale=0.31]{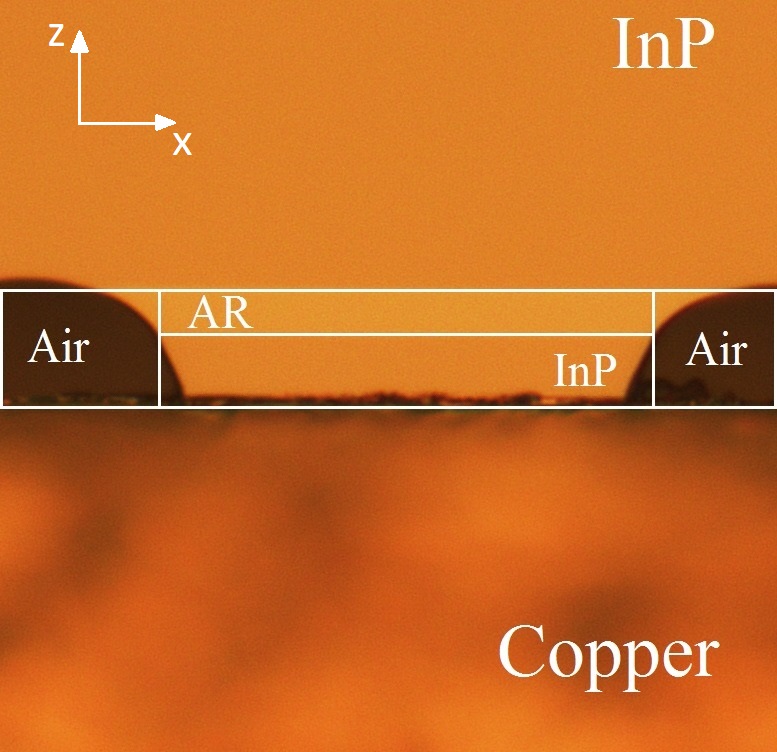}
\caption{The real image of a QCL structure with schematic sketch, indicating the simplifications made for numerical modelling. In the latter the ARn is considered as uniform media with some averaged thermal parameters. For this calculation ARn properties are taken similar to InAs all but \textcolor{black}{thermal conductivity.}}
\label{fig-struct}
\end{figure}

To reproduce the thermal behaviour of QCL structure we employ the time dependent two-dimensional heat equation.

\begin{equation}
    c \rho \frac{\partial T (x, z, t)}{\partial t} - \kappa ( \frac{\partial^2}{\partial x^2}+\frac{\partial^2}{\partial z^2})T(x, z, t)  = q_v (x, z, t)
\end{equation}

\noindent where $c(x,z)$, $\rho(x,z)$ and $\kappa(x,z)$ are spatial functions, characterising heat capacity, density and thermal conductivity of a QCL design, $T(x, z, t)$ and $q_v (x, z, t)$ are time dependent two dimensional functions of temperature distribution and volumetric heat sources.
The spatial model providing relevant description of heat transfer processes is depicted in Fig.~\ref{fig-struct}. We use a two dimensional map with domains mimicking thermal properties of InP-claddings, copper, used as a heat sink, and ARn correspondingly (see Table~\ref{tbl:heparam}). Here we consider the direct modelling of the ARn heterostructure unnecessary for our goals, so for simplicity the ARn heat capacity and density are adopted equal to properties of InAs. 
These parameters vary for +/- 10\%  for any compound of III-V materials  \cite{adachi92}, thus this approximation appears to be valid. The only unknown parameter of the system is an ARn thermal conductivity, which significantly and unpredictably decrease in solid solution comprising fluctuating chemical formula\cite{maycock67,jaffe19} and plenty of interfaces\cite{sood14, mei15}. Thus we simply adjust it to the experimental data.

\begin{table}
\caption{Material parameters, which are used to perform heat equation modeling for QCL}
\begin{tabular}{|m{1.3cm}|m{2.2cm}|m{1.2cm}|m{3cm}|} \hline
material & heat capacity, J~(g~K)$^{-1}$ & density, g~cm$^{-3}$ & thermal conductivity, W~(cm~K)$^{-1}$ \\ \hline \hline
InP    & 0.31  & 4.8  & 0.68 \\
copper & 0.40  & 8.9  & 4.00 \\
ARn     & 0.31  & 5.5  & n/a  \\ \hline
\end{tabular}
\label{tbl:heparam}
\end{table}

\begin{figure}
\includegraphics[scale=0.31]{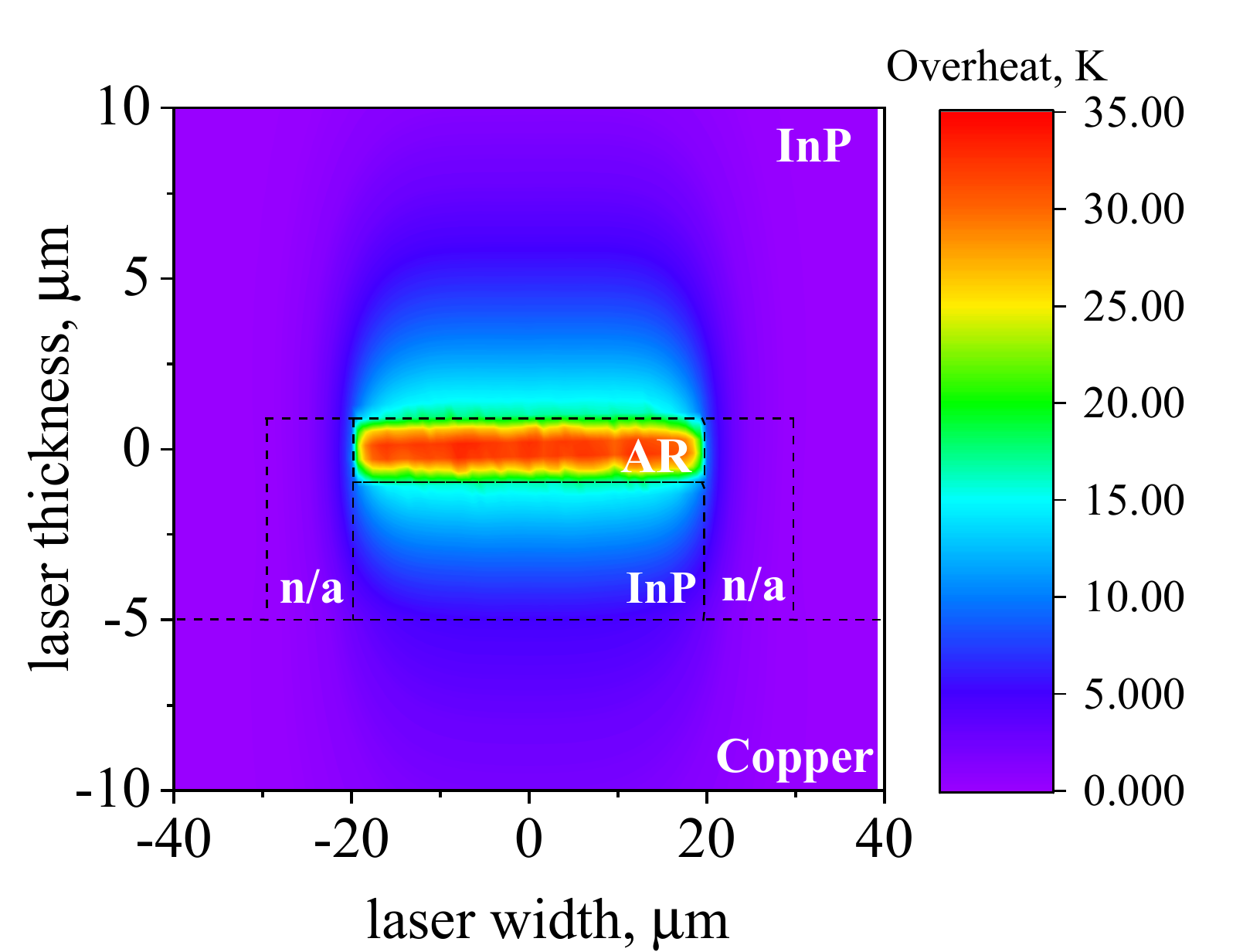}
\caption{Heatmap of QCL cross section after 500~ns 6.5~A pump pulse. The scale of the heat map is adopted from data forming Fig.~\ref{fig-struct}}
\label{fig-heatmap3d}
\end{figure}

As was mentioned above the expected ARn overheat is about 25~K. 
In order to obtain such a value during modeling the effective ARn thermal conductivity was calibrated and made equal to 0.07~W~(cm~K)$^{-1}$, which is 10~\% of the value inherent to InP. The heat distribution map after 500~ns of 6.5~A pump pulse is depicted in Fig.~\ref{fig-heatmap3d}. The color idicates the heatmap in the structure taken from Fig.~\ref{fig-struct}. Zero coordinates indicate the middle point of the two-dimensional AR. It can be clearly seen that the energy released in a QCL during operation is confined in the vicinity of an AR. The part of heat dissipated out from an ARn is approximately a half of the released amount. In the end of 500~ns pulse the typical diffusion length is only 5~$\mu$m, showing that the system is far form steady state regime when thermal resistance formalism is valid. To achieve a pure steady state regime the ARn should heat up the area as large as about 50~$\mu$m with a diffusivity constant of InP.


\begin{figure}
\includegraphics[scale=0.31]{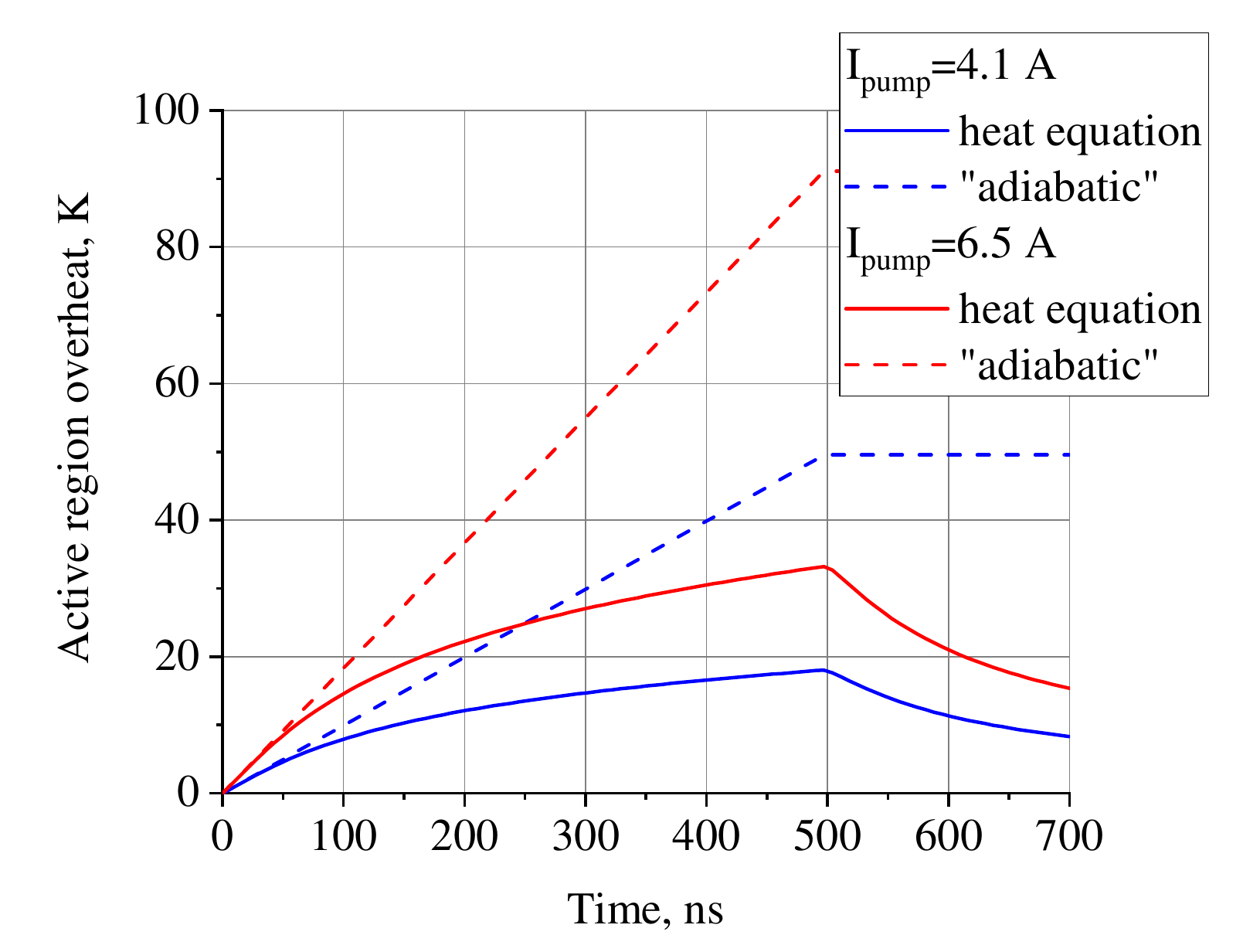}
\caption{Temperature profile of an ARn center under operation with two specific values of pump current. The pump pulse length is 500 ns. The dashed lines represent the adiabatic estimations, when all the heat is assumed to be enclosed in an AR. \textcolor{black}{One can see that during the first 100 ns and especially in the beginning of the pulse the dashed and solid lines show very little difference, proving that under experimental conditions there is quite low thermal dissipation. }}
\label{fig-tempprofile}
\end{figure}

The temperature profile of the ARn center is depicted in Fig.~\ref{fig-tempprofile}. One can find that in the very beginning of a pulse (about 100~ns) the energy release can be considered as adiabatic. So here we can estimate that to the end of modelled pulse real temperature is twice higher than the one in adiabatic approximation.

\section{discussion}

\subsection{ARn thermal conductivity evaluation}

As it follows from our modelling results the effective thermal conductivity of InGaAs/AlInAs ARn in perpendicular direction is about 0.05~W(m$\cdot$K)$^{-1}$, which is 5-10 times\cite{sood14} lower than this parameter for common pure bulk materials such as GaAs (0.55~W(cm$\cdot$K)$^{-1}$), InAs (0.27~W(cm$\cdot$K)$^{-1}$), InP (0.68~W(cm$\cdot$K)$^{-1}$). Possible reasons for such a drastic decrease may lie in the two main reasons: the decrease of heat transfer due to the interfaces properties and the thermal properties of chemical solution. The proper consideration of interface properties, for example acoustic mismatch model (AMM) and diffuse mismatch models (DMM), allows one to conclude that when a layer thickness is larger than a phonon mean free path than inter-layer heat transfer does not changes significantly. \cite{mei15, sood14, jaffe19, chen97}. The main factor limiting in plane thermal conductivity is a quality of interface (most likely roughness, which is root-mean-square effective interface height), obviously the specular-like interface induce more effective heat conductivity inside the layer compared to rough one allowing the back-scattering of the phonon. Also a layer with diffusive (rough) interface lowers the transfer when the thickness becomes smaller. However, this effect also does not give much contribution to the heat-transfer suppression. As a result, any addition of interfaces characterizing by different fabrication technologies\cite{jaffe19} reduces this parameter by co-factor of 1.05 to 2.5. So the Kapitza resistance between alloy layers has an upper bound of 0.1 m$^2$K~GW$^{-1}$ and is negligible compared to the intrinsic alloy resistances\cite{sood14}, even for 2~nm thick layers. 

Thus we consider the alloy properties as the main factor of the heat transfer drop. The experimental research unambiguously demonstrated\cite{jaffe19} that the thermal conductivity of solid In$_{0.5}$Ga$_{0.5}$As and In$_{0.5}$Al$_{0.5}$As alloys are of 0.02-0.04~W(cm$\cdot$K)$^{-1}$. Earlier studies\cite{maycock67,goryunova68} support the idea that the thermal conductivity of III-V solid solutions mostly decreases due to isovalent substitutions in the lattice.

\subsection{Fundamental limit on the efficiency of heat transfer from AR: analytical evaluation}

As it was highlighted in the previous sections an ARn thermal conductivity is characterized by the lowest value compared to other pure materials used. This means that an ARn itself is a bottleneck for effective thermal management of QCL. As it follows from our own data and literature review the main contribution to the thermal conductivity degradation is provided by the use of solid solution (alloy) of III-V materials. The secondary effects are due to interfaces in the structure of ARn and heat spreader configuration.

\begin{table}
\caption{Thermal conductivity degradation rate and ratio of its minimum and maximum values specific for compounds induced by alloying of III-V materials\cite{goryunova68}.}
\begin{tabular}{|m{2.2cm}|m{3.5cm}|m{2.2cm}|} \hline
alloy & $d \kappa/ d x $, W(cm$\cdot$K$\cdot$mol)$^{-1}$ &  $\kappa_\text{min}/\kappa_\text{max}$, \% \\ \hline \hline
InAs$\rightarrow$GaAs  & 2.0 & 18 \\
GaAs$\rightarrow$InAs  & 1.8 & 22 \\ \hline
GaP$\rightarrow$GaAs   & 5.6 & 11 \\ \hline
InAs$\rightarrow$InP   & 0.8 & 30 \\ 
InP$\rightarrow$InAs   & 4.2 & 14 \\ \hline
\end{tabular}
\label{tbl-slopecomparis}
\end{table}

\begin{figure}
a)\includegraphics[scale=0.3]{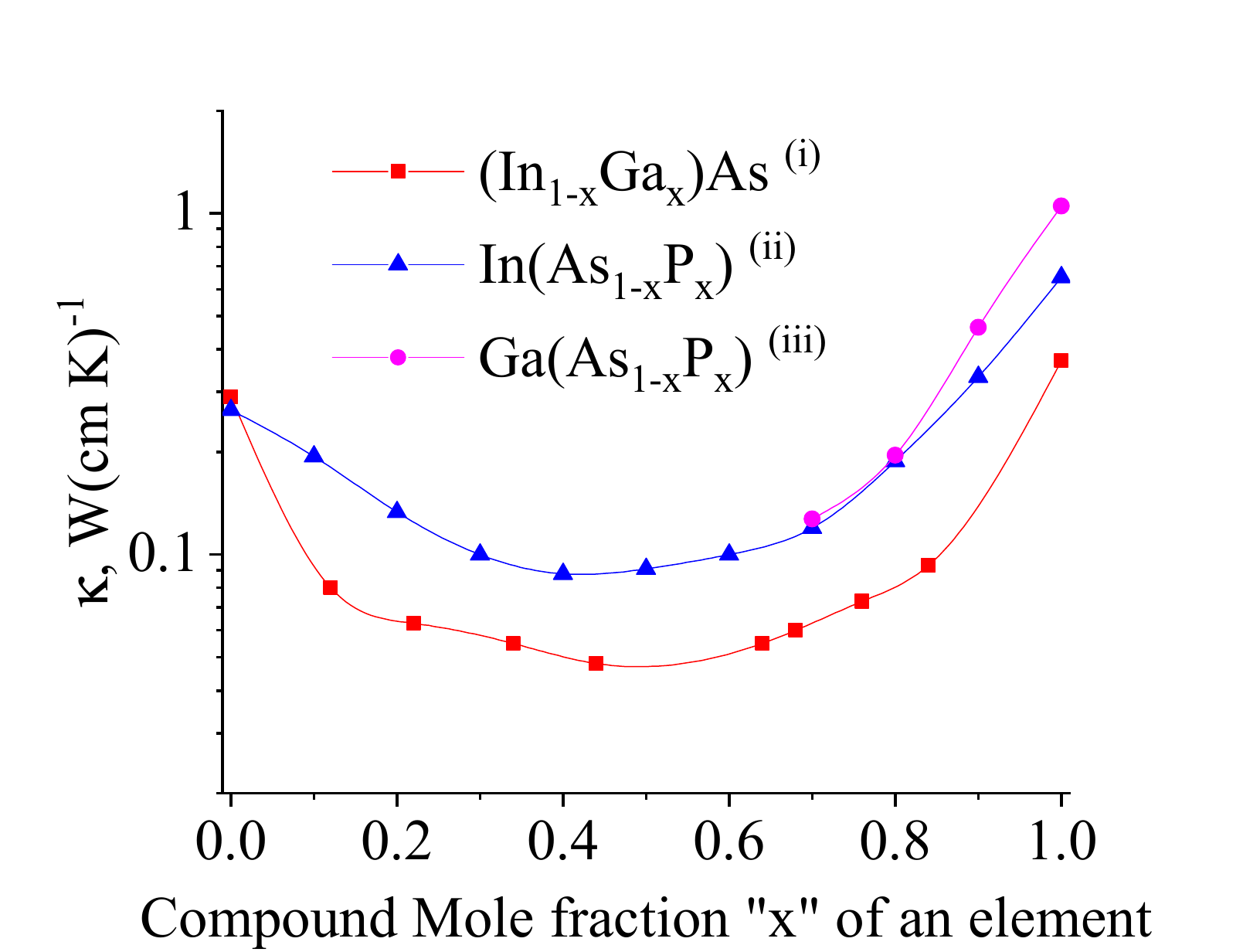}
b)\includegraphics[scale=0.3]{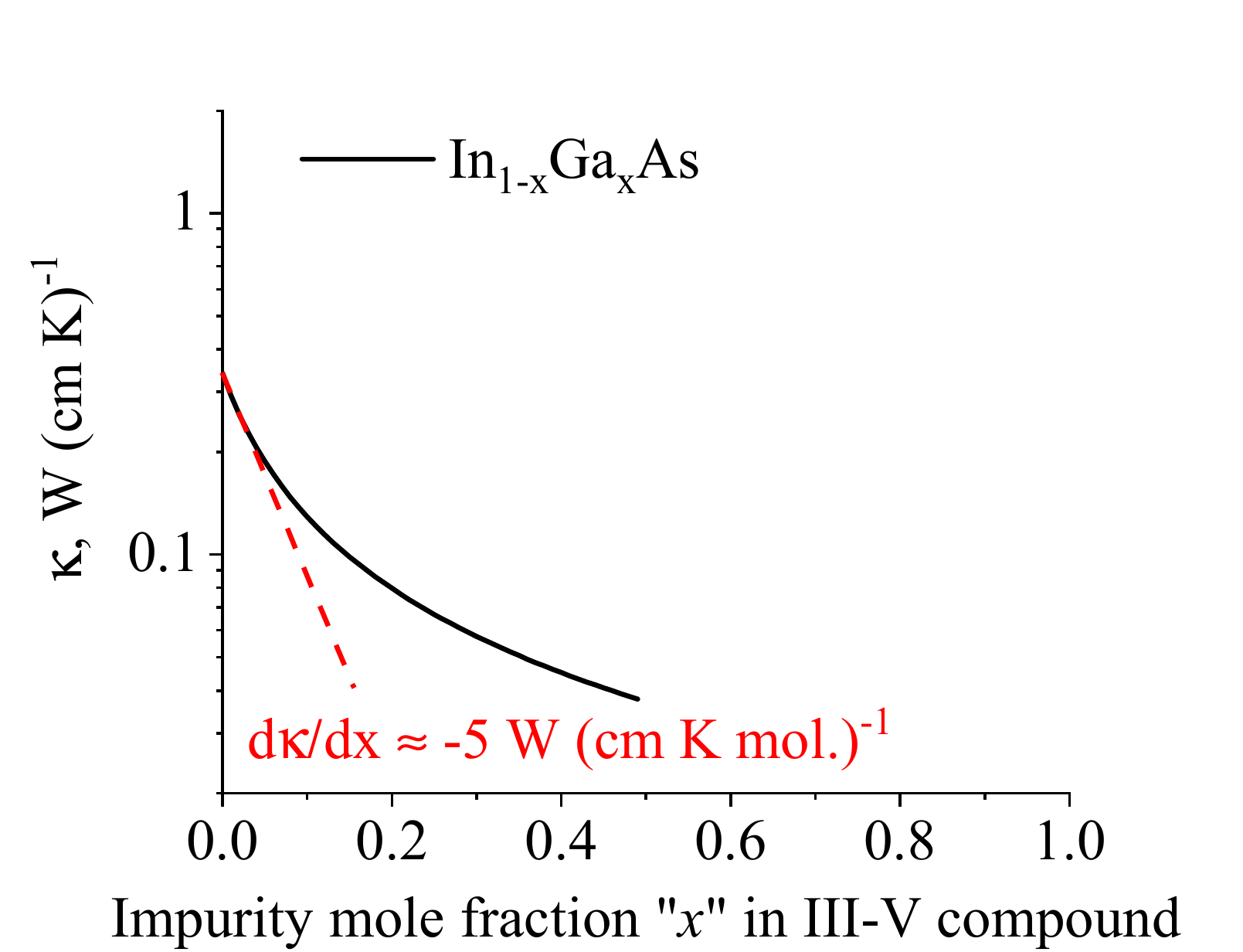}
\caption{ a) The log-scale dependence of the thermal conductivity on the III-V compounds' chemical formula adopted from Ref.~\onlinecite{abrahams59} -- ${(i)}$, Ref.~\onlinecite{maycock67}-- ${(ii)}$, Ref.~\onlinecite{wiess59} -- ${(iii)}$, initially collected by Goryunova\cite{goryunova68}. One can easily see that clear compound shows better thermal characteristics than any mixed solution.  b) The log-scale  thermal conductivity degradation of In$_{1-x}$Ga$_x$As calculated according to eq.~\ref{eq-kappamicro} is depicted in solid black line. The dashed red line indicates the rate of thermal conductivity degradation in the low isovalent substitution concentration limit. One can find the theoretical estimations are quite close to the experimental results.}
\label{fig-thermconddegrad}
\end{figure}

Now we consider the physical aspects of the main challenge of QCL thermal management in the simple semi-analytical manner. The pure bulk III-V materials manifest the value of thermal conductivity of the order of 0.5~W(cm$\cdot$K)$^{-1}$ at room temperature. On the one hand this value is temperature dependent, on the other hand it is rather stable with respect to the trace amount of impurities in the raw materials\cite{maycock67}. This allows us to conclude, that in a pure binary III-V alloy the main effect limiting heat dissipation is a phonon-phonon scattering.

Assume that the macroscopic heat transfer 
is an envelop function characterized by microscopic process of phonon diffusion:

\begin{equation}
L_{diff}=\sqrt{2 \alpha  t}
\label{eq-macrodiff}
\end{equation}

\noindent where $\alpha = \frac{\kappa}{\rho c}$ [cm$^2$s$^{-1}$] is a thermal diffusivity. At the same time the diffusion length of individual phonon with time $t$ [s] can be assessed via the following formula:

\begin{equation}
L_{diff}=\sqrt{\lambda \nu t}
\label{eq-microdiff}
\end{equation}

\noindent where $\lambda$ is a mean free path of the phonon [cm], $\nu$ is a speed of sound [cm$\cdot$s$^{-1}$]. Basing on the initial assumption we equate heat diffusion lengths derived via macro- and microscopic approaches 
and obtain the intrinsic (natural) mean free path for effective phonon mode in pure InAs compound as:

\begin{equation}
\lambda_{nat} = \frac{2 \alpha}{\nu} \approx \text{(for InAs)} \approx\frac{2*0.2}{4\cdot10^5}=10^{-6}~\text{cm}.
\end{equation}

If a pure semiconductor is diluted with impurity ions to prepare solid solution (alloy) the contaminating ions reduce the mean free path according to the following assessment:

\begin{multline}
\frac{1}{\lambda_{tot}}=\frac{1}{\lambda_{nat}}+\frac{1}{\lambda_{imp}}=\frac{1}{\lambda_{nat}}+\sigma_{imp}N_{imp}=\\=\frac{1}{\lambda_{nat}}+\sigma_{imp}N_0x
\end{multline}

\noindent where $\sigma_{imp}$ [cm$^2$] is a cross section of phonon-impurity scattering, $N_{imp}$ [cm$^{-3}$] is a concentration of unit cells accommodating impurity ions, $x$ is a ratio in the chemical compound formula (e.g. In$_{1-x}$Ga$_x$As), $N_{0}$  [cm$^{-3}$] is a concentration of unit cells in initial lattice (4.4$\cdot$10$^{21}$~cm$^{-3}$ for InAs). 

Rewriting eqs.~\ref{eq-macrodiff}~and~\ref{eq-microdiff} one can also get the following expression for the thermal conductivity:

\begin{equation}
\kappa= \frac{ c \rho \nu }{2} \frac{1}{\frac{1}{\lambda_{nat}}+\sigma_{imp}N_0x }
\label{eq-kappamicro}
\end{equation}

\noindent Thus the degradation rate of pure InAs thermal conductivity with contamination in the low concentration limit reads:

\begin{equation}
\frac{d \kappa}{ d x } =  \frac{ c \rho \nu }{2} \frac{-\sigma_{imp}N_0}{(\frac{1}{\lambda_{nat}}+\sigma_{imp}N_0x )^2}
\label{eq-kappaslopemicro}
\end{equation}

The only value to guess in the eqs.~\ref{eq-kappamicro}~and~\ref{eq-kappaslopemicro} is the cross-section characterizing phonon-impurity scattering $\sigma_{imp}$. We assume that the value of the cross-section equals the square of the lattice constant ($\sigma_{imp}$=3.6$\cdot$10$^{-15}$~cm$^2$). The thermal conductivity degradation profile and its slope in the limit of low impurity concentration for the guessed value are depicted in Fig.~\ref{fig-thermconddegrad}. The plotted profile is typical for III-V compounds (see e.g. Refs.~\onlinecite{maycock67, goryunova68, jaffe19}). The comparison of the theoretically estimated slope for the low concentration limit and experimentally assessed values collected in Ref.~\onlinecite{goryunova68} are presented in Table~\ref{tbl-slopecomparis}.

\subsection{CW regime and structure safety}

\begin{figure}
\includegraphics[scale=0.31]{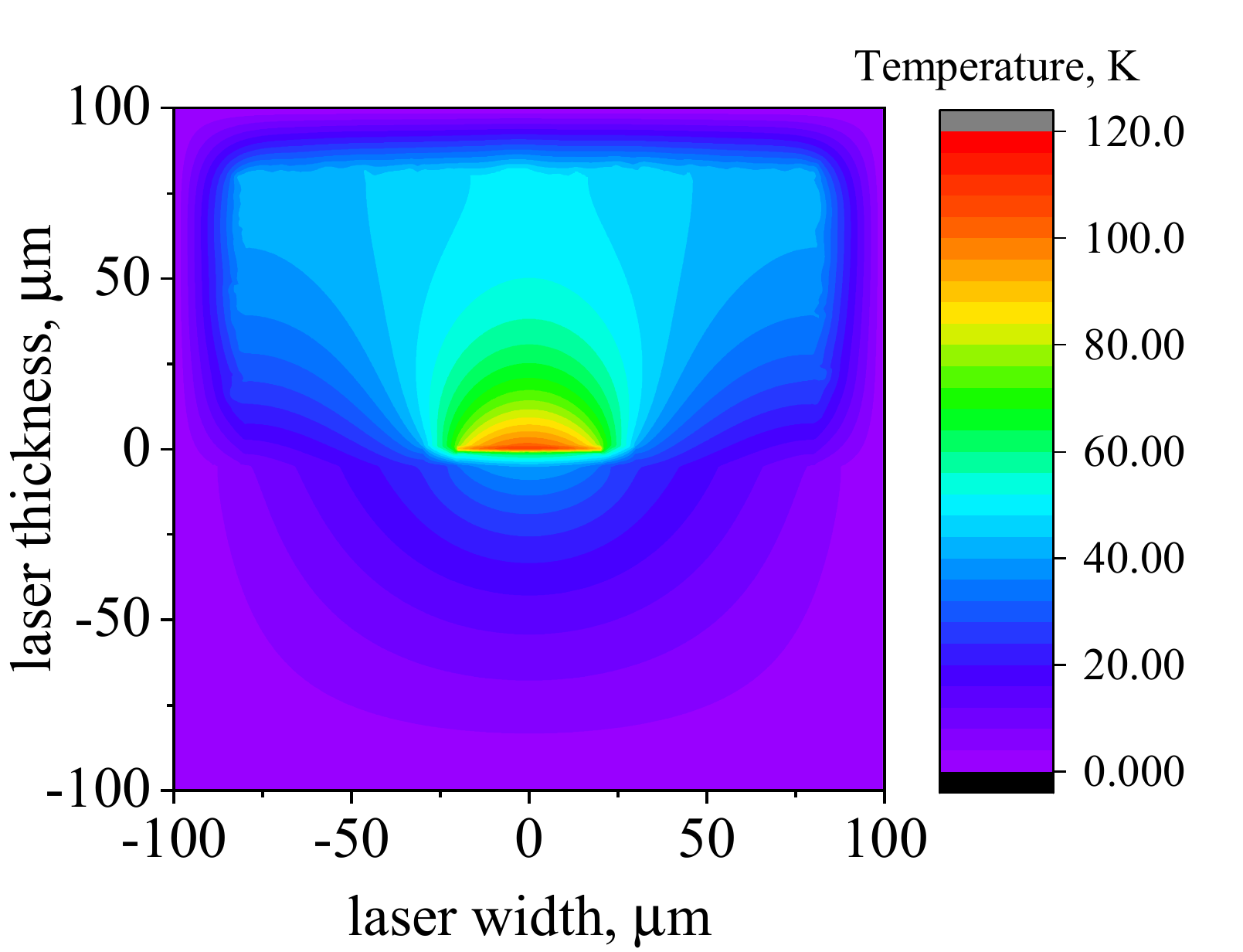}
\includegraphics[scale=0.31]{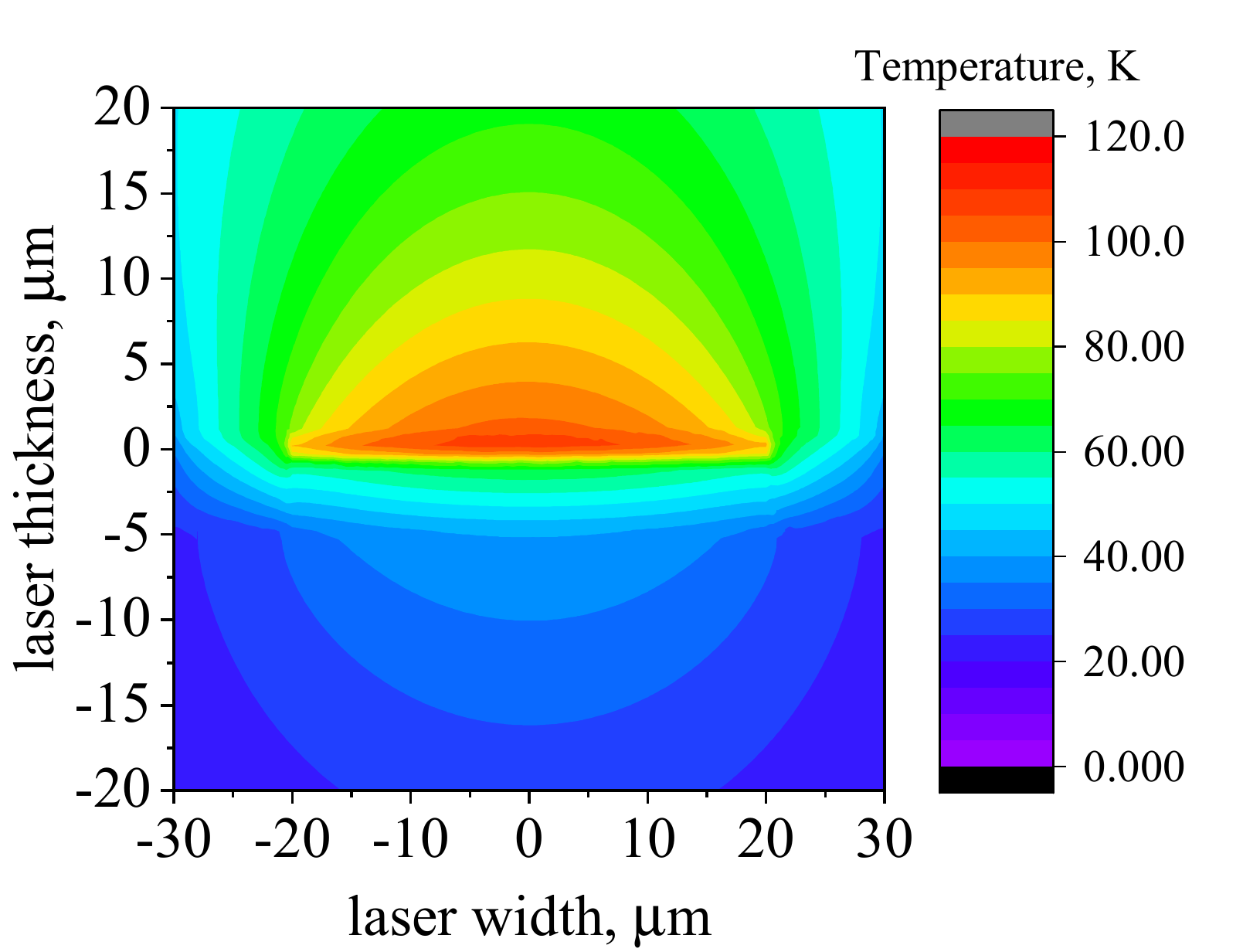}
\caption{
Heat map of QCL device schematically similar to the one in Fig.2 in CW regime under pump current of 6.5~A. The copper level starts at -5$\mu$m-level. The size of a sample for this calculation is 0.2*0.2~mm. The upper panel shows the reached steady state operational mode, showing the isoterms of heat spreading.  Lower panel is a zoomed ARn from the upper panel.
}
\label{fig-cwregime}
\end{figure}

 Fig.~\ref{fig-exp} clearly shows that a laser operating in high power regime does not allow one to reach pulse duration over units of microsecond. However, with the calibrated value of an ARn thermal conductivity one may assess the ability to reach continuous wave regime for the mentioned structure. Here we neglect the technical details of heat transfer with the coolant, assuming that there is a reservoir supporting room temperature on the boundaries of the modelled box with the given size (see upper panel in Fig.~\ref{fig-cwregime}). Adopting these simplifications we run the simulation for period long enough and control if the system reaches steady state. Calculations show, that after about 10~$\mu$s of steady state pump the temperature of a central part of an ARn stabilizes, with an overheat of about 100~K (see lower panel of Fig.~\ref{fig-cwregime}).

Fig.~\ref{fig-cwregime} shows numerical modelling of temperature distribution for the real QCL structure in the steady state. This can be generalized by introducing equivalent thermal resistance scheme comprising comprehensible primitives. To proceed we make the following assumptions:
\begin{itemize}
    \item assume that the device operates at room temperature (RT);
    \item assume that the temperature on the outer surface of the heat spreader is somehow stabilized equal to RT;
    \item neglect a distribution of the temperature inside an AR;
    \item neglect the heat leakage from the sides of the laser ridge;
    \item geometry of an ARn is needle-like, meaning that its linear dimensions in the \textcolor{black}{XZ plane}(facet plane) are negligible compared to the copper dimensions (See Fig~\ref{fig-struct}). This allows us to consider cylindrical distribution of isotherms in the the copper spreader. The last can be easily seen in the lower panel of the Fig~\ref{fig-cwregime}.
\end{itemize}
Under these assumptions the corresponding serially connected primitives are presented in  Fig.~\ref{fig-thermresscheme}

In a real device the ARn may be separated from the coolant by the two thermal resistors: InP rod and copper cylinder\cite{zhang10}, which are coupled in the series circuit.The thermal resistance of the first part is:

\begin{equation}
T_{InP}\approx T_{1D}=\frac{P}{\kappa L w} h= \frac{0.0004}{0.68*0.3*0.0040} P \approx 0.5 [\frac{\text{K}}{W}] P
\end{equation}

The second component of heat spreader is read as:

\begin{equation}
T_{Cu}\approx T_{rad}=2\frac{P}{\kappa \pi L } Ln(\frac{R}{\rho})= 2\frac{Ln(\frac{0.1}{0.0040})}{4*3.14*0.3}   P \approx 2 [\frac{\text{K}}{W}] P
\end{equation}
The details for these expressions can be found in the Appendix~\ref{app:Tres}


It can be seen in Fig.~\ref{fig-cwregime} that the overheat of upper and lower sides of 4~$\mu$m InP rod is about 70~K and 50~K, which results in almost linear 20~K temperature gradient. The remaining overheat of 50~K is distributed over copper heat spreader. The ratio of these two thermal differences 20~K/50~K is close to  the ratio of assessed thermal resistances  0.5~$\frac{\text{K}}{\text{W}}$ / 2~$\frac{\text{K}}{\text{W}}$.


In the estimations above we did not consider the ARn thermal properties. Apparently the thermal conductivity is of the order of magnitude lower than that value for pure InP. Taking into account the 2~$\mu$m thickness of an ARn this creates an additional temperature increase in the center of QCL and significant non-uniformity of a temperature distribution over QCL volume. Such drastic localized thermal variations create strong lattice strains, therefore it should be a parameter for monitoring under operation condition to prevent potential device damage\cite{sin15, zhang10}. 

\section{conclusion}
In this work we discuss the fundamental limits of the QCL performance, related to the Joule heating. We measured the intensity degradation in InGaAs/InAlAs QCL and find it following a trend typical for adiabatic heat release in the very beginning of the pump pulse. The heat equation modeling reveals that the leakage of Joule heat released in ARn starts approximately after 100~ns of the pump pulse beginning, which is related to the creation of temperature gradient in the vicinity of AR. The process of radiator heating takes more than tens of microseconds and finishes when the steady state thermal distribution is reached. The latter corresponds to a QCL CW mode and allows rather simple description by thermal resistance approximation.

In both pulsed and CW regimes the cornerstone of the effective thermal management is a thermal conductivity of an ARn which suffers (i.e. decreases) from inevitable presence of interfaces in the QCL structure and usage of III-V solid solutions. The mentioned aspects of laser design are due to an effective moderation of phonon diffusion by roughness of the layers interface and presence of scattering isovalent substitutions in the alloy lattice. The overall effect for thermal conductivity reduction estimated based on measurements is an order of magnitude, i.e we assess the effective value of 0.07~W~(cm~K)$^{-1}$. The low effective thermal conductivity of an ARn leads to the high non-uniformity of the temperature distribution which creates additional strains in the device media.

\section{Acknowledgements}
Authors acknowledge funding from from the Russian Science Foundation (project 21-72-30020)

\appendix


\section{Thermal resistance of heat spreader primitives}\label{app:Tres}

\begin{figure}
\includegraphics[scale=0.31]{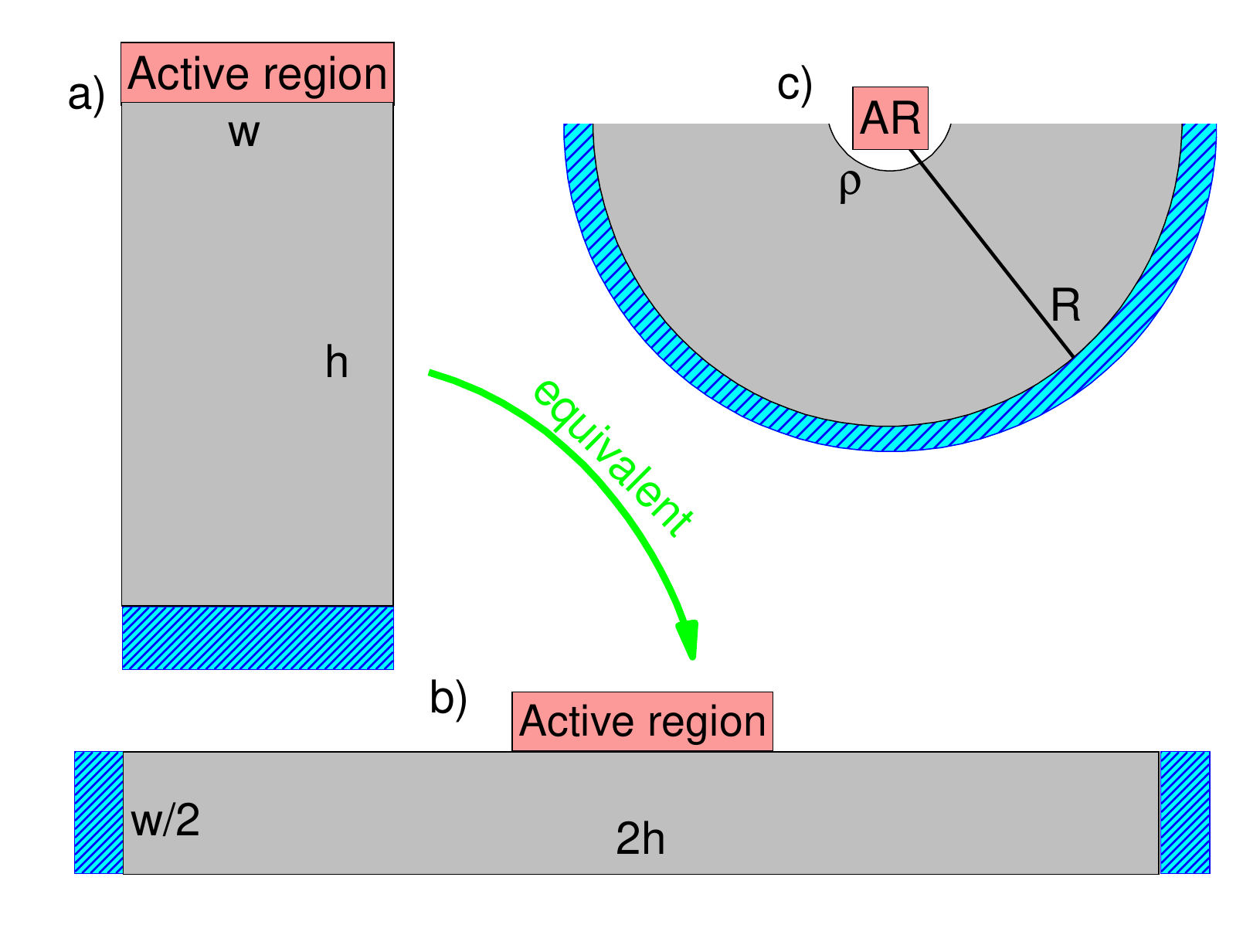}
\caption{Scheme of flat (a and b) and cylindrical (c) heat spreaders. Red bars denote AR, blue bars denote the contact of the heat spreader with temperature ``reservour'', grey bars depict heat spreader medium.}
\label{fig-thermresscheme}
\end{figure}

Consider the heat equation in the volume of planar and cylindrical heat spreaders (see Fig.~\ref{fig-thermresscheme}).

\begin{equation}
\Laplace T =0 
\end{equation}

The first model (a) and b) panels in Fig.~\ref{fig-thermresscheme}) is a rod like heat spreader with the height ($h$) significantly larger than the width ($w$). The non-trivial solution of the heat equation in a such geometry is a linear function:

\begin{equation}
T_{1D} (x)= k x+ b
\end{equation}

Calibration of the parameters in the previous equation using boundary conditions result in the following maximum power heat spreader is able to evacuate keeping the overheat of $\Delta T$:

\begin{equation}
T_{1D} (x)=\frac{P}{\kappa L w} (h-x)
\end{equation}

The radial symmetry of the heat spreader leads to modification of the temperature distribution in the following form:

\begin{equation}
T_{rad} (r)= \frac{P}{\kappa L \pi} Ln(\frac{R}{r})
\end{equation}



\section{Other samples}
\begin{figure*}
\includegraphics[scale=0.25]{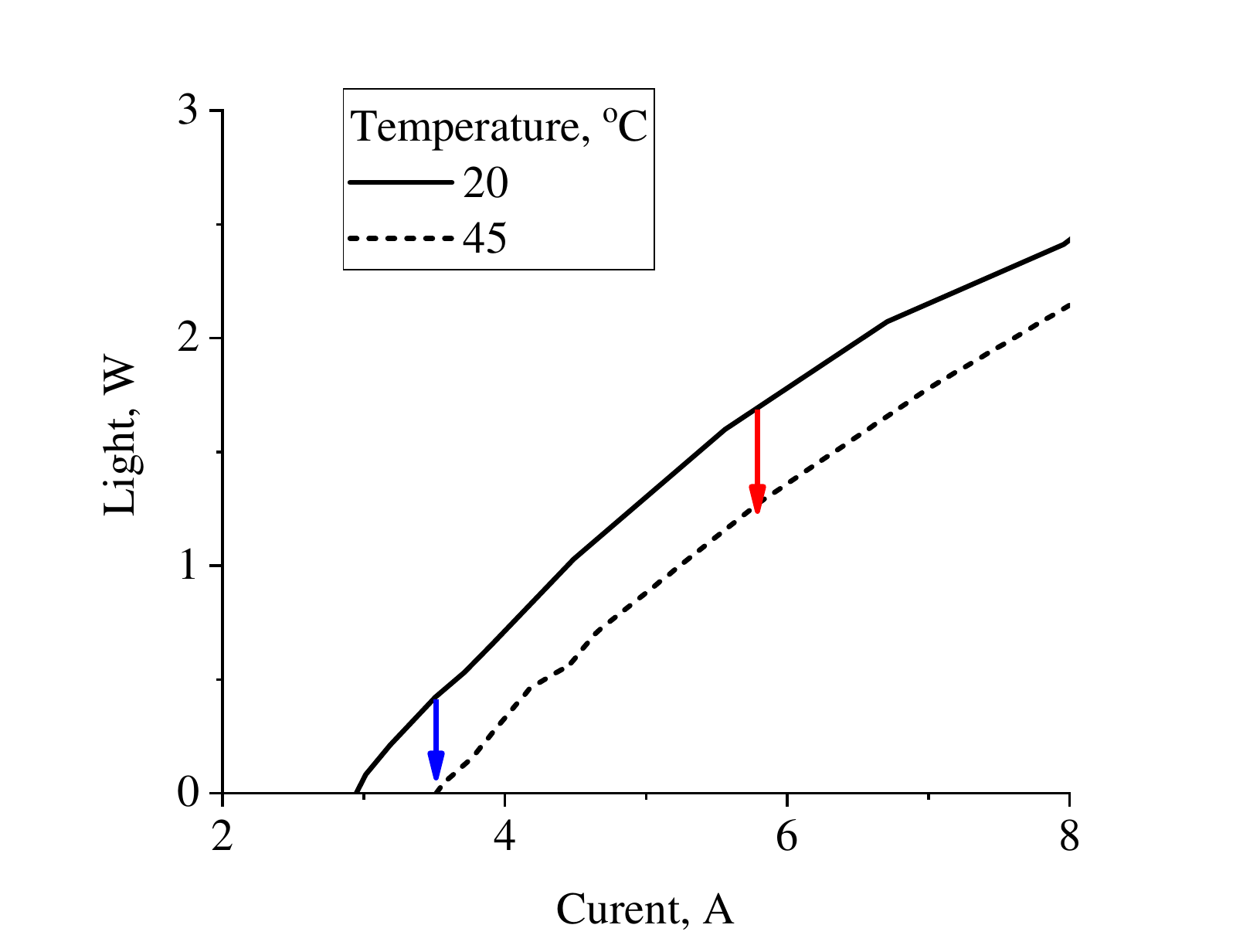}
\includegraphics[scale=0.25]{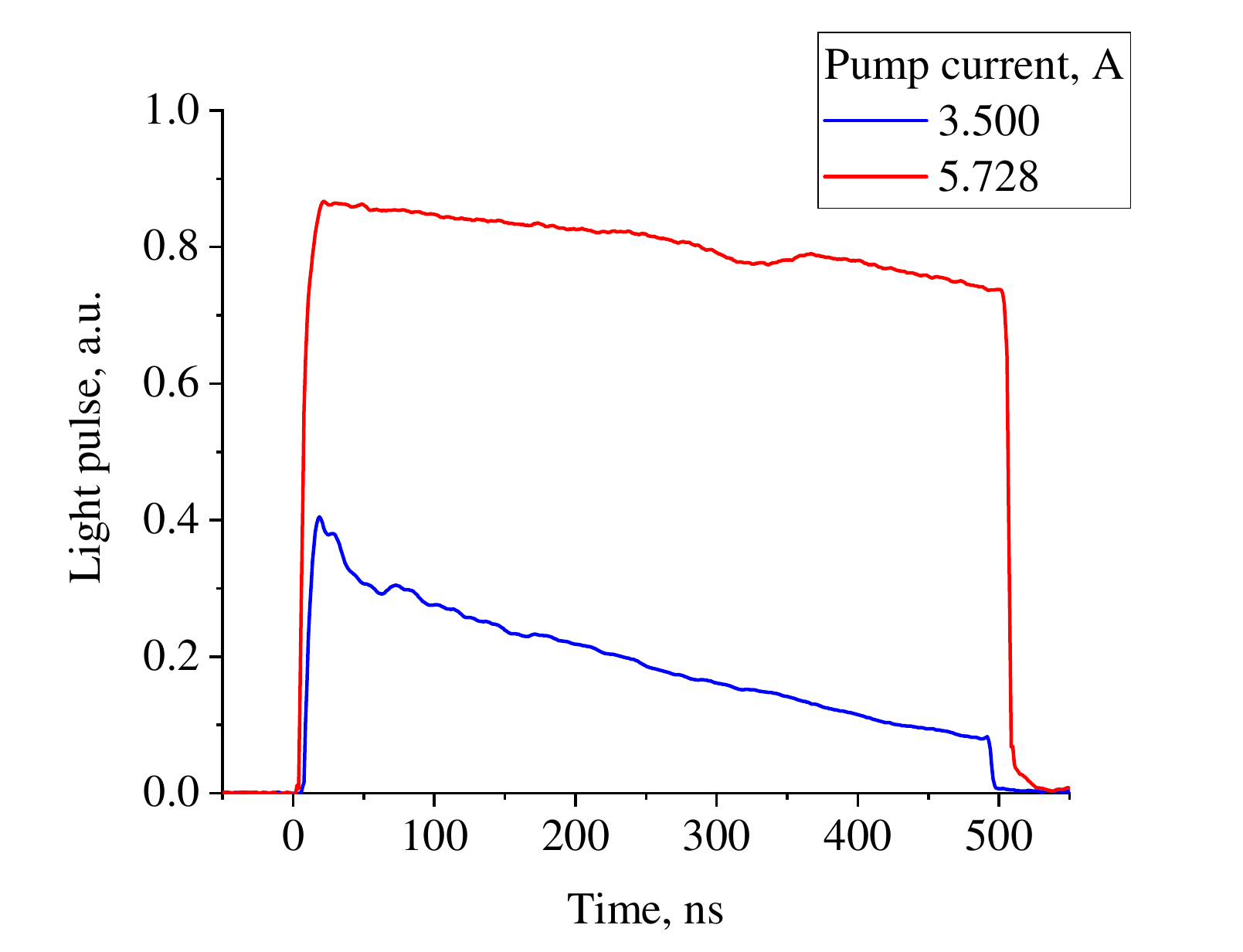}
\includegraphics[scale=0.25]{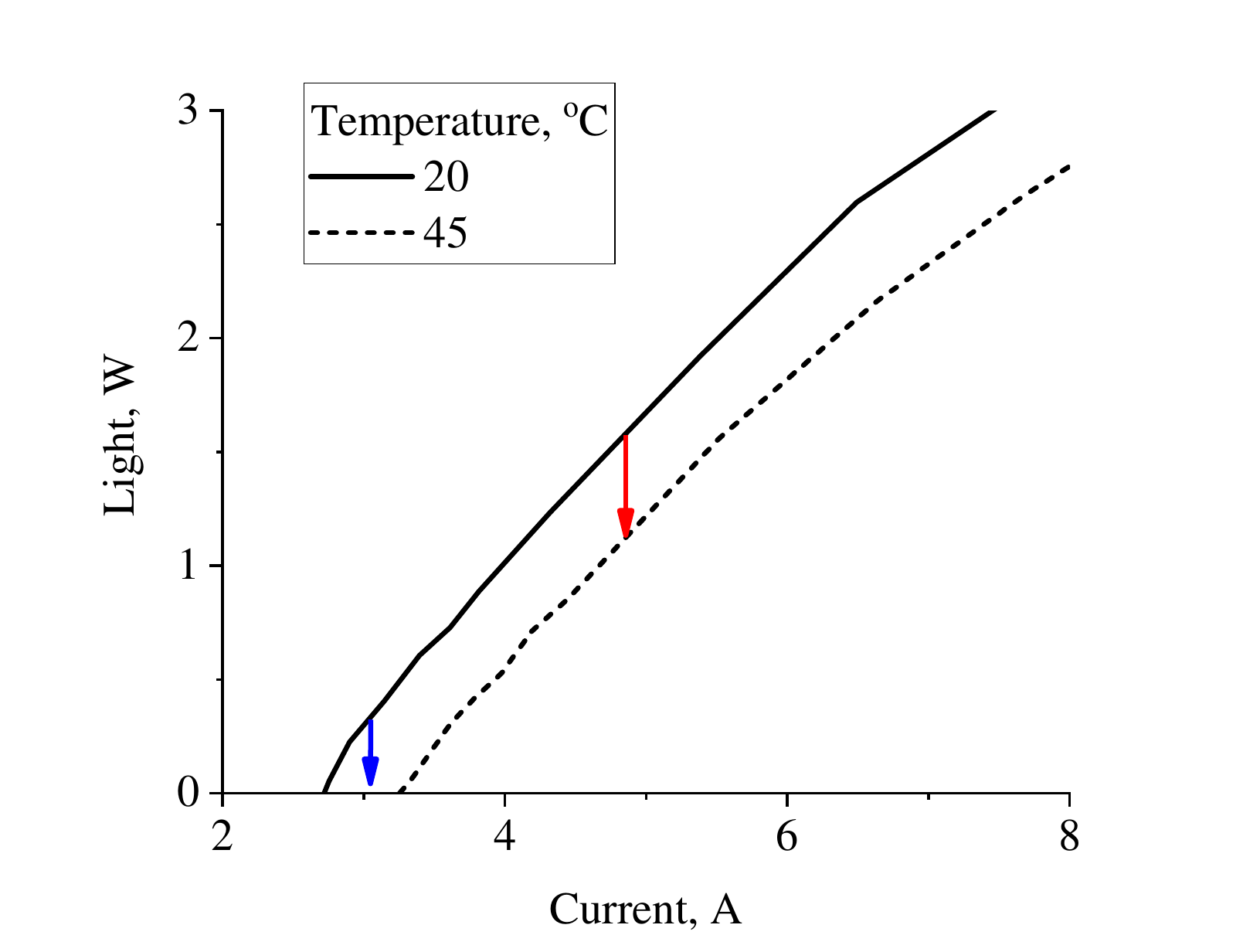}
\includegraphics[scale=0.25]{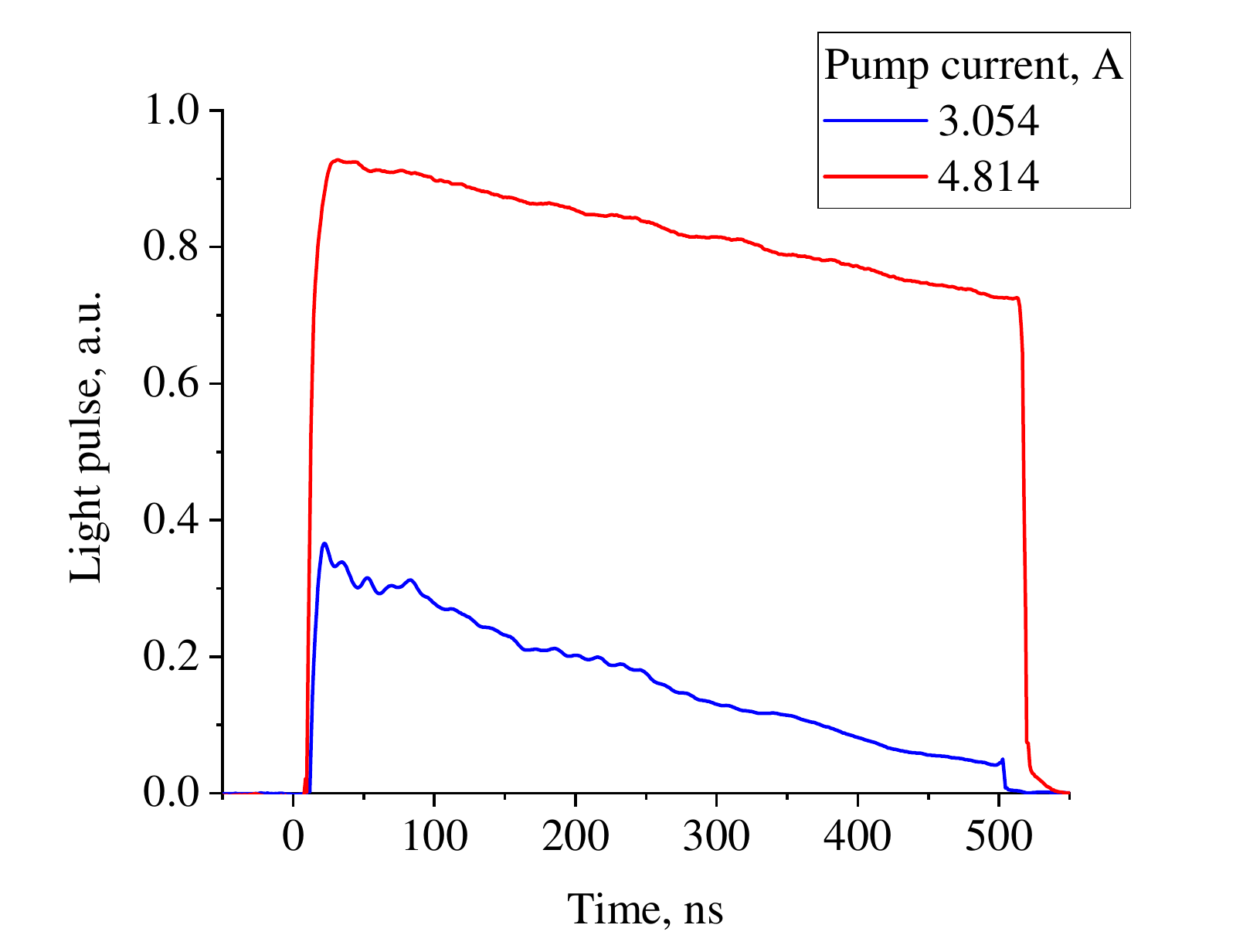}
\includegraphics[scale=0.25]{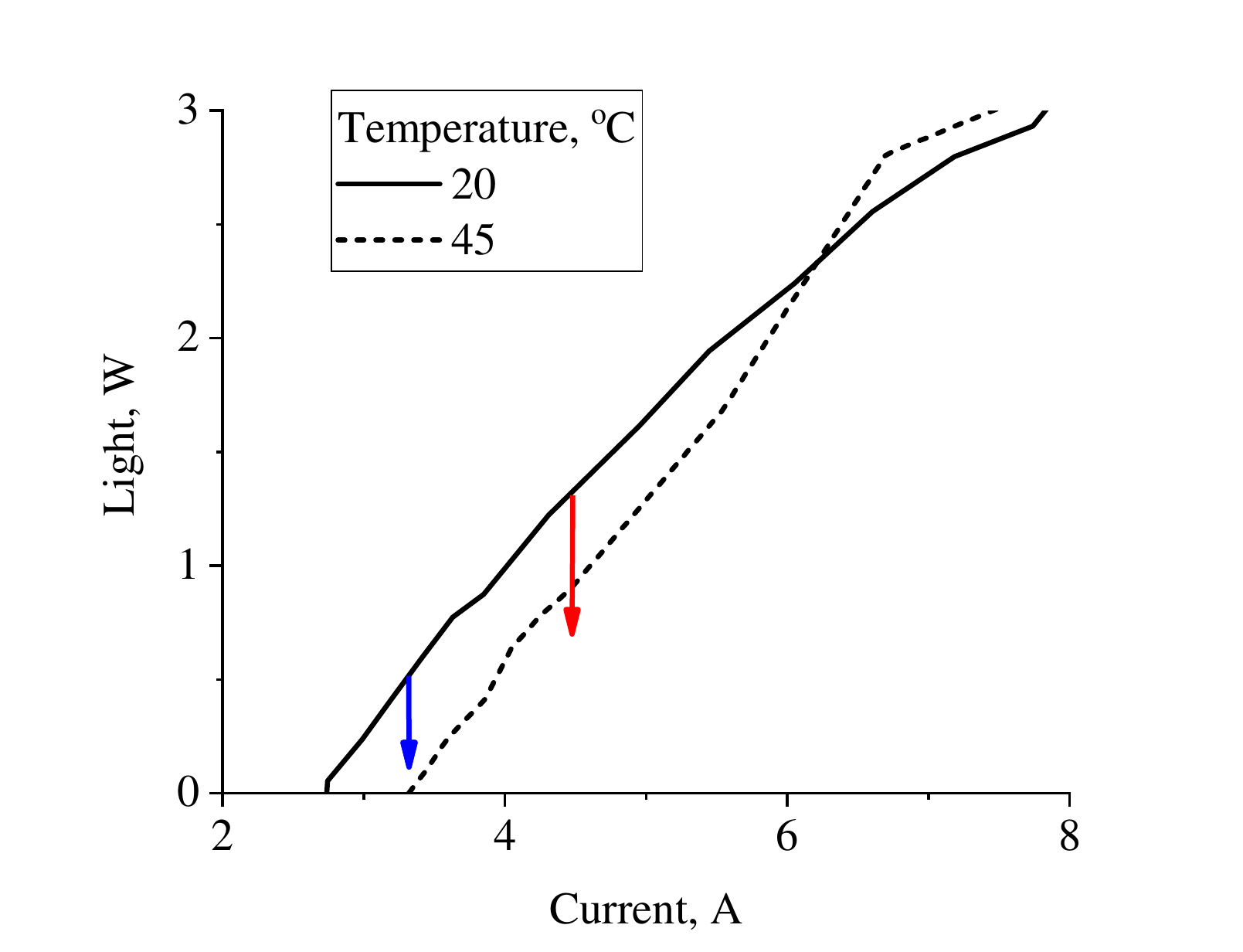}
\includegraphics[scale=0.25]{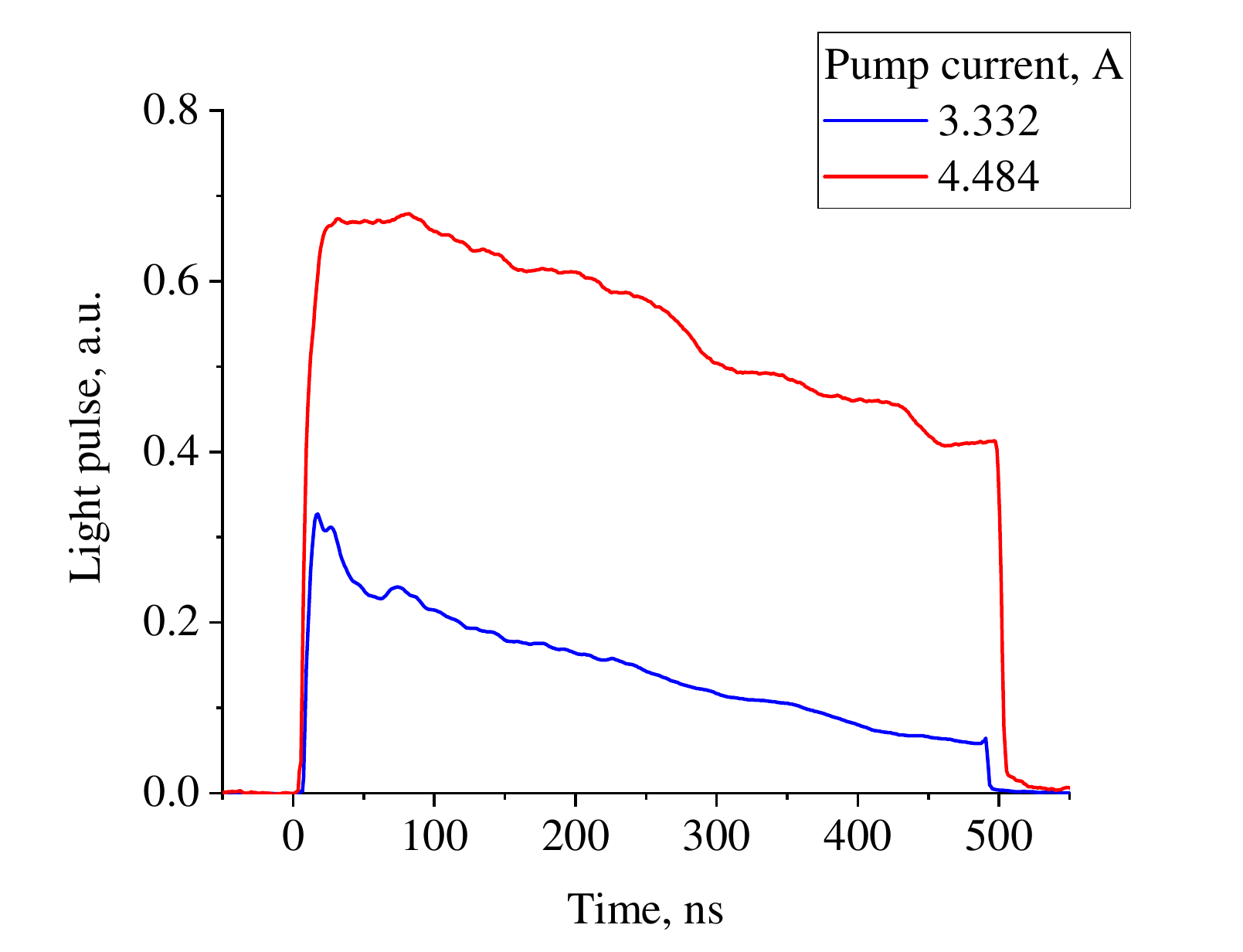}
\caption{Left column - light current characteristics of several samples of InGaAs/InAlAs QCLs. Right column - time dependent light pulses of the mentioned lasers under specified values of pump current. The degradation of the light intensity with time is also depicted in the left column using vertical arrows having corresponding color.}
\label{fig-exp-app}
\end{figure*}
\clearpage
\newpage

\bibliography{bibliography}

\end{document}